\documentclass[aps,prd,preprint,groupedaddress]{revtex4-1}
\usepackage[utf8]{inputenc}
\usepackage{amsmath, amsfonts, mathtext, amssymb, mathrsfs,subfigure,wrapfig}
\input epsf
\usepackage[utf8]{inputenc}
\usepackage{graphicx} 

\usepackage{textcomp}
\newcommand*{\No}{\textnumero} 

\begin{document}

\title{Horndeski/Galileon black hole shadows}

\author{D.A. Tretyakova}
\email[]{daria.tretiakova@urfu.ru}
\affiliation{Institute for Natural Sciences, Ural Federal University, Lenin av. 51, Yekaterinburg, 620083, Russia}

\author{T.M. Adyev}
\affiliation{Institute for Natural Sciences, Ural Federal University, Lenin av. 51, Yekaterinburg, 620083, Russia}

\date{\today}

\begin{abstract}
We examine observational signatures of the asymptotically de Sitter branch of static spherically symmetric Horndeski/Galileon  solutions. Our analysis  reveales that possible deviations of the  photon sphere size from those of the GR Schwarzshild solution are vanishing compared to the resolution of modern radio-interferometric observations. These deviations are heavily suppressed by the bounds coming from the weak-field observation.  We conclude that shadow size would not be a useful characteristic to distinguish the Horndeski/Galileon static spherically symmetric spacetimes  in the foreseeable future.
\end{abstract}


\maketitle

\section{Introduction}

Since it's foundation General Relativity is constantly subject to experimental tests. Such tests are still topical for multiple alternative gravity theories that emerged due to cosmological \cite{Bamba:2012cp} and theoretical reasons \cite{Woodard:2009ns}. 
However most of the observations available nowadays probe the weak field regime \cite{Turyshev:2008dr}. Strong field tests however would allow to constrain (ore, may be, detect) alternative gravity much better. Such tests can be conducted by means of different methods. For example, by analyzing the relativistic spectra profiles, originating from the innermost regions of the black hole accretion disc, we obtain the information on the innermost stable circular orbit \cite{Tanaka:1995en}. The ultimate distinguishing feature of the black hole is, however, the event horizon. To a distant observer, the event horizon casts a relatively large "shadow" due to the bending of light by the black hole, and this shadow is nearly independent of the black hole spin or orientation. The predicted size ($\sim 50 \mu as$ \cite{PhysRevD.90.062007}) of this shadow for Sgr A* approaches the resolution of current radio interferometers. Sizes, shapes of shadows  and strong-field gravitational lensing for different types of black holes were calculated
in multiple papers, for example \cite{1538-4357-528-1-L13,PhysRevD.87.044057}.

Doeleman et al. \cite{Doeleman:2008qh} evaluated the size of the smallest spot for the black hole at the Galactic Center with the VLBI technique in mm-band. Theoretical studies showed that the size of the smallest spot near a black hole practically coincides with the shadow size because the spot is the envelope of the shadow  \cite{PhysRevD.80.024042, 1538-4357-528-1-L13}. Thus studying the smallest spot provides useful information about the spacetime properties. Paper \cite{Doeleman:2008qh} claimes that the intrinsic diameter of Sgr A* is $37^{+16}_{-10} \mu as$ at the $3\sigma$ confidence level. So a Schwarzschild black hole is just
marginally consistent with observations. In extended gravity the shadow size and shape differ: they are affected by the corrections put into the Schwarzshild metric by the additional degrees of freedom. Hence measuring the size and shape of the black hole may help to evaluate parameters of black hole metric and put alternative gravity to test. 

In this paper we consider scalar-tensor gravity, a widely accepted alternative to the General Relativity. The most general scalar-tensor  action resulting in the second order field equations was  proposed by Horndeski \cite{Horndeski:1974wa}. The same result was rediscovered by studying the covariant Galileons \cite{Deffayet:2009mn, Deffayet:2009wt},  a ghost free scalar effective field theory containing higher derivative terms that are protected by the Galileon shift symmetry. The action for the  Horndeski/Galileons scalar-tensor model  reads
\begin{equation}
S = \int dx^4 \sqrt{-g} \left( \zeta R - \eta \left( \partial \phi \right)^2 + \beta G^{\mu\nu} \partial_\mu \phi \partial_\nu \phi - 2\Lambda \right) \label{ac},
\end{equation}
here $ G^{\mu\nu} $ is the Einstein tensor, $ \phi $ is the scalar field,  $ \zeta>0, \eta$ and $\beta$ are model parameters. This model is a part of the so-called Fab Four action, and the term, containing the Einstein tensor is usually denoted as John. Though the Horndeski non-trivial kinetic coupling sector is not exhausted by this action we will restrict our consideration to the action above, since most of the static spherically symmetric solutions we are about to consider relate to this framework\footnote{Note also that the second component of the non-trivial kinetic coupling sector, the so-called Paul term, experiences problems with  describing  neutron
stars \cite{Maselli:2016gxk}.}.  
The model \eqref{ac} is known to admit a rich spectrum of cosmological solutions (see \cite{Starobinsky:2016kua} and references therein) describing the late-time acceleration and the inflationary phase. Moreover, for $\eta\neq 0$ it admits solutions for which the $\Lambda$-term is totally screened, while the metric is not flat but rather de Sitter with the Hubble rate proportional to $\eta/\beta$. This offers an exciting opportunity to describe the late time cosmic self-acceleration while screening the vacuum $\Lambda$-term and hence circumventing the cosmological constant problem. Henceforth it would be interesting to pursue the study of the solutions of \eqref{ac} at the astrophysical  and solar system scale.

This model can be integrated completely in the static and spherically symmetric sector \cite{Charmousis:2015aya} and numerous black hole solutions are presented in the literature. The key ingredient of this solutions is a scalar field linear time dependence, which seems like a natural feature on a cosmological background. Although the Vainstein screening mechanism is generally at work in Horndeski gravity, in the case of a minimal coupling of the scalar field to matter no screening radius can be posed. The solution, considered as a candidate to represent astrophysical objects must then display de Sitter asymptotic\footnote{The asymptotic structure of the solutions might also be interesting in the context of AdS/CFT correspondence and in brane cosmology, along the lines of \cite{Birmingham:2001dq}.}. As we explain below, the framework \eqref{ac} contains a whole class of such solutions, so we need to find some selection mechanisms, ore, at least, bound on the parameter space. A way to see if this spacetimes may represent viable astronomical objects. This might help in choosing viable solutions and parameter ranges, reducing the ambiguity related to the vacuum solution choice. 

 The purpose of this paper is thus to threat the local  spherically symmetric solutions of  \eqref{ac}  as astronomical objects and see if any signatures, distinguishing this spacetimes from the Schwarzschild one, can be detected in the near future. To do so we study the null-geodesics describing light rays propagation near the compact object. 

The paper is organized as follows. In section \ref{s1} we briefly summarize the properties of the black hole-like solutions for the action \eqref{ac}. Section \ref{s2} contains solution's parameter estimates which are essential for the analytical consideration. We proceed with analyzing null geodesics in section \ref{s3}. Conclusions are given thereafter.

\section{Horndeski/Galileon black hole} \label{s1}
Spherically-symmetric static black hole solutions  were given in several papers \cite{Rinaldi:2012vy}-\nocite{Babichev:2013cya, Charmousis:2015aya, Babichev:2015rva, Anabalon:2013oea}\cite{Minamitsuji:2013ura}
\begin{equation}
ds^2 = h(r)dt^2 - \frac{dr^2}{f(r)} - r^2d \Omega^2
\label{ds^2}
\end{equation}
These solutions are constructed by means of the following equations: 
\begin{eqnarray}
&& f(r)  =  \dfrac{( \beta + \eta r^2) h(r)}{\beta \left( rh(r)\right)'},\label{f0} \\
&& h(r)  =  -\frac{\mu}{r} + \frac{1}{r} \int\dfrac{k(r)}{\left( \beta + \eta r^2\right)}dr, \label{f00}\\
&&\phi(r)  =  qt+\psi(r), 
\end{eqnarray}
where $ \mu$ plays the role of the mass term and $k$ is derived by  the constraint equation:
\begin{eqnarray}
&& q^2\beta\left( \beta + \eta r^2\right)^2 -
\left( 2\zeta\beta + \left( 2\zeta\eta - \lambda\right) r^2\right)k + C_0 k^{\frac{3}{2}} =0. \label{k}
\end{eqnarray}
Here $C_0$ is an integration constant. The key ingredient here is a mild linear time of the scalar field. It helps to evade the scalar field  derivatives singularity on the event horizon  \cite{Charmousis:2014zaa} and nullifies the no-hair theorem at the same time. The Galileon symmetry is keeping the field equations  consistent with the static ansatz and provides asymptotically flat or de-Sitter solutions. 

Equations \eqref{f0},\eqref{f00},\eqref{k} provide a huge variety of solutions  for different parameter combinations of $C_0$ and q. Among those possessing realistic asymptotic behavior is the stealth de Sitter one 
\begin{eqnarray}
f(r) & = & h(r)=1-\cfrac{\mu}{r} +Ar^2 , \label{f1}\\
A&=&\cfrac{\eta}{3\beta}, q^2=(\zeta\eta+\beta\Lambda)/(\beta\eta), C_0=(\zeta\eta-\beta\Lambda)\sqrt{\beta}/\eta.
\end{eqnarray}
Curiously this solution completely evades the vacuum cosmological constant $\Lambda$. The authors of \cite{Charmousis:2015aya} prove that \eqref{f1} is not an isolated solution, instead it may be extended to a full branch of de Sitter like black hole solutions of a kind
\begin{eqnarray}
h(r) & = & C-\frac{\mu}{r} +Ar^2 +\Delta, \qquad \Delta=B\dfrac{\arctan (r \gamma)}{r \gamma}, \label{h_gen}
\end{eqnarray}
which is asymptotically de Sitter for $A<0$.  Solutions can be constructed along the guidelines presented in \cite{Charmousis:2015aya}. The authors support this statement by obtaining a solution, deviating slightly from the de Sitter one due to the small parameter $\epsilon$
\begin{eqnarray}
A & = & -\cfrac{\eta}{3|\beta|}, B= \cfrac{2(1+\gamma^2)\epsilon}{\zeta+y}, \epsilon <<|y-1|, \nonumber\\
\gamma&=& \sqrt{\cfrac{\eta}{|\beta|}}\cfrac{\zeta+y}{\zeta-3y}, C=1-\cfrac{2\epsilon}{\zeta+y}, y=\cfrac{\Lambda|\beta|}{\eta}, \label{apds}
\end{eqnarray}
By examining the known solutions from \cite{Rinaldi:2012vy}-\cite{Minamitsuji:2013ura} we see that \eqref{h_gen} is a very common expression, joining most of the solutions (thought the differ by the valyes of $A,B,C,\gamma$) and many others of a kind can be constructed due to master equations. So a class of solutions governed by \eqref{h_gen} is vast and not limited to asymptotically de Sitter spacetimes. Given a class of vacuum solutions instead of a single solution we a re forced to find some algorithms allowing us to choose those being acceptable as a real-life object candidates. To fill this need we are about to study the observational properties of the asymptotically de Sitter black hole solutions, namely the event horizon shadow.  So in what follows we will use the metric in the form \eqref{h_gen}, assuming that the results apply to any solution of that kind. 

\section{Parameter estimates} \label{s2}


We now briefly review studies of the static spherically symmetric solutions from the non-mininal derivative coupling sector of Horndeski theory, presented in the literature. 
Firstly one should keep in mind that for the action \eqref{ac} the scalar also acquires dynamics via the Jonh term. Therefore the condition for the solution to be ghost-free does not just boil down to the ``right'' kinetic term sign. See e.g. \cite{Deffayet:2010qz}, where a Galileon is demonstrated to acquire stable  cosmological solutions for the unusual sign of the standard kinetic term.


Minamitsuji \cite{Minamitsuji:2014hha} investigated the stability of BH solutions under massless scalar perturbations in the nonminimal derivative coupling subclass. The quasinormal modes can be computed, and no unstable modes were found. Considering the same BH solutions, Cisterna et al. \cite{Cisterna:2015uya} found that these black holes are stable under odd-parity gravitational perturbations as well. 
The stability conditions of hairy black holes in the non-minimal derivative coupling sector can be extracted from \cite{Ogawa:2015pea}, keeping in mind that  $X \neq Const$ due to the radial dependence of the scalar field.

Any modification of GR must be consistent with  astrophysical and Solar  System  scale constraints,  which  are  very  stringent. The Schwarzschild solution in GR also describes the exterior of any spherically symmetric body in the weak field limit (hence the Solar System), and so must do it's analog in extended gravity theory. 
So now we  turn to the black hole spacetime in question and consider weak-field observations to pose some bounds on the correction terms. One of the well-studied gravitational effects is the frequency shift for the satellite on the Earth orbit
\begin{equation}
\cfrac{\delta \nu}{\nu}=1-\sqrt{\cfrac{h(R+d)}{h(R)}}\approx \cfrac{V(R+d)-V(R)}{c^2},
\end{equation}
up to the first order in the weak field approximation with $d$ being the satellite orbit height and $R$ - the Earth surface radius, $V$ - the corresponding gravitational potential. For the metric \eqref{h_gen} there should be an additional shift, related to the deviation of h(r) from the Schwarzshild solution
\begin{eqnarray}
&& 2\cfrac{\delta \nu}{\nu}\approx \delta_{Schw}+\delta_1+\delta_2 =\nonumber \\
&& \qquad \left ( \cfrac{\mu}{R}-\cfrac{\mu}{R+d} \right)+ A( (R+d)^2-R^2) + \left ( \Delta(R+d)-\Delta(R) \right).
\end{eqnarray}
Current frequency measurements show no deviation from GR, hence we can make bounds using the frequency measurement accuracy $10^{-14}$ achieved in the GP–A redshift experiment\footnote{$d= 15\times 10^{3} km, M_{\oplus}=5,972\times 10^{24} kg, R_{\oplus}=6371 km, \\ G_0=6.67384 \times 10^{-11} m^3 kg^{-1} c{-2}$} \cite{PhysRevLett.45.2081}. Numerical estimates show that $\delta_1$  does not exceed the accuracy of the relative frequency measurement when
\begin{eqnarray}
A&<&2.4\times 10^{-29}m^{-2}.
\end{eqnarray}
This bond allows us to set $A\approx 0$ when considering accretion and null geodesics for example for the Sgr A* black hole or any smaller one since the conditions $Ar^2<<\mu/r$ and $Ar^2<<\Delta$ are well satisfied  within the corresponding  $100r_{Schw}$ distance\footnote{$M_{Sgr A}=4\times 10^6 M_{\odot}$}.  So, we may neglect the de Sitter term on astronomical scales and the metric can be considered in the form
\begin{eqnarray}
h(r) & =& C-\frac{\mu}{r} +\Delta \label{h0}.
\end{eqnarray}
This does also agree with the fact that the expansion of the universe does not manifest itself in the Solar System up the the current measurement accuracy.  We can further use the equation \eqref{f0} to define $f$ with respect to $\eqref{h_gen}$ and neglect terms $\sim Ar^2$ to keep our approximation. This would imply
\begin{equation}
f(r)\approx\cfrac{h(r)}{B+C} \label{fgm}. 
\end{equation}

To keep the frequency shift  in agreement with the experiment $\delta_2$ should also be adjusted to the experimental error. This  can be done by both $B$ and $\gamma$ and it is unclear what bound can be placed here. However we can carry out some interesting conclusions by analyzing concrete solutions. 
Take the approximately Schwarzshild - de Sitter solution \eqref{apds} as an example. The solution is valid for $y/\zeta\in (-1,1/3)$. The second multiplier in $\gamma$ is of order unity except the special case $\zeta\approx 3y$. Hence, excluding this fine tuned case we can see that $\gamma \sim \sqrt{|A|}$. So we can further suggest that due to the  smallness of $\gamma$, $\arctan(r\gamma)\approx r\gamma$ and hence
\begin{eqnarray}
\Delta&\approx&B, \\
h(r) & =& (C+B)-\frac{\mu}{r}. \label{h1}
\end{eqnarray}
This would be the spacetime of a black hole with a global monopole (up to the difference in f(r)).  This kind of black hole  was previously studied  in the literature for flat and de Sitter spacetime, revealing the expressions for the deflection angle \cite{Shi:2009nz}, perihelion precession \cite{Hao:2003rz} and accretion disc radiant energy flux \cite{1674-1056-23-6-060401}. Note that the \eqref{h1} metric does not display the solid angle deficit since the unity is restored in $f(r)$ due to \eqref{fgm} and $h(r)$ is ambiguous by the constant time rescaling.  

To support the statement that the conclusions withdrawn from the structure of \eqref{h_gen} are of general nature, consider the solution from \cite{Babichev:2013cya}:
\begin{eqnarray}
C&=&1 , A =  \dfrac{\gamma^2}{3}\dfrac{\zeta - y}{3\zeta + y} ,  B =  \dfrac{(\zeta+y)^2}{4 \zeta^2 - (\zeta+y)^2}, \gamma = \sqrt{\eta /3\beta},  \label{bab}
\end{eqnarray}
where  $\gamma \sim \sqrt{A}$ analogously to the case above.

We will further use \eqref{h0} as well as \eqref{h1} to study null geodesics and withdraw some conclusions.

\section{Null geodesics} \label{s3}
The motion of the photons  around a black hole  is governed by the geodesic equations
\begin{eqnarray} \label{geodesic}
&& \left( \frac{du}{d\varphi} \right)^2 =  f(u)\left[\cfrac{E^2}{h(u)} \cfrac{}{L^2}  -u^{2}\right] \\
\end{eqnarray}
with $u=r^{-1}$ being the inverse radius. Taking \eqref{fgm} and \eqref{h1} into account we obtain for the global monopole case
\begin{eqnarray} 
&& (C+B)\left( \frac{du}{d\varphi} \right)^2 = \frac{E^2}{L^2} -  (C+B) u^2 + \mu u^3 = R(u)\\
\end{eqnarray}
 For the cubic $R(u)$ the circular orbit of photons corresponds to the case of two equal real roots, which would be the case if
\begin{eqnarray}
&& R(u) = 0, \quad R'(u) = 0.
\end{eqnarray}
The corresponding roots are 
\begin{eqnarray}
&&u_1 = u_2 = \dfrac{2  (C+B)}{3 \mu}, u_3 = - \dfrac{ (C+B)}{3 \mu}, \\
 &&\dfrac{E^2}{L^2} = \dfrac{4 A^3}{27 \mu^2}.
 \end{eqnarray}
This would result in the following radial geodesic equation:
\begin{eqnarray}
&&\left( \frac{du}{d\varphi} \right)^2 =  \cfrac{\mu}{(C+B)} \left( u - \dfrac{2  (C+B)}{3 \mu}\right) ^2 \left( u  + \frac{ (C+B)}{3 \mu}\right). 
\end{eqnarray}
where we get after the integration
\begin{eqnarray}
&&u = - \frac{ (C+B)}{3 \mu} + \frac{ (C+B)}{ \mu} \tanh^2 \left( \frac{\varphi - \varphi_0}{2}\right), 
\end{eqnarray}
with $ \varphi_{0} $ being an integration constant.
The radius of the photon sphere  can be found from the above by taking the limit $ \varphi \to \infty $:
\begin{eqnarray}
&&r_s = u_s^{-1} = \frac{3 \mu}{2  (C+B)}. \label{ps}
\end{eqnarray}
Here we read-off the obvious requirement $(C+B)>0$. The Schwarzschild value would be achieved by $(C+B)=1$. 

Interesting to note, that $Ar^2$ we neglected would not contribute to the above anyway. The null geodesic equation \eqref{geodesic} could be rewritten as
\begin{eqnarray}
&& \dot{r}+ \mathcal{V}(r) =  \frac{E^2}{2  (C+B)}, \\
&&\mathcal{V}(r) = f(r)\cfrac{L^2}{r^2},
\end{eqnarray}
where dot denotes the derivative with respect to $\tau$, some affine parameter along the geodesic. Circular orbits are defined by the stationary points of the potential $\mathcal{V}(r)$. But $Ar^2$ would cancel out with $L/r^2$ to a constant and make no contribution to the derivative of $\mathcal{V}(r)$.

The equation \eqref{ps} generally implies that the photon sphere radius can be altered drastically depending on the particular values of B and C.  At this point we would like to recall the paper \cite{Hao:2003rz} that contains the the light deflection angle formula for the global monopole, which would allow us to pose some bounds on $(C+B)$. The  deflection angle in our notation would be
\begin{equation}
\delta \varphi\approx  \cfrac{2\mu}{(C+B)^{3/2}R_0}.
\end{equation}
Given the current measurement accuracy we could put bounds on $(C+B)$. The paper \cite{10875} based on a number of VLBI measurements of angular separations of strong quasistellar radio sources passing very close to the Sun, claims a good agreement of the observations with GR. The reported relative error in the  deflection angle is  $6.2\times 10^{-4}$, which would imply
\begin{equation}
|1-(C+B)|<3\times 10^{-4}. \label{b0}
\end{equation}
Given the  constraint above we see that $r_s$ may deviate from the Schwarzschild value by no more than $10^{-4}$. Detecting such a difference would nowadays be impossible since the resolution is of order of the shadow size itself.


\section{Discussion and conclusions}
In this paper we considered observational signatures of the asymptotically de Sitter branch of static spherically symmetric Horndeski/Galileon  solutions. Our analysis  revealed that possible deviations of the  photon sphere size from those of the GR Schwarzshild solution are vanishing compared to the resolution of modern radio-interferometric observations. These deviations are heavily suppressed by the bounds coming from the weak-field observation. 
This we conclude that shadow size would not be a useful characteristic to distinguish the Horndeski/Galileon static spherically symmetric spacetimes  in the foreseeable future. More realistic indicators would include S-stars motion and accretion disc radiant energy flux observations, as is was found in \cite{Tretyakova:2016knb}.

\section{Acknowledgements}
This work was supported by Russian Foundation for Basic Research via grant RFBR \No 16-02-00682.

\bibliographystyle{unsrt}
\bibliography{mybib}

\end{document}